\documentclass[aps,prl,twocolumn,showpacs,groupedaddress]{revtex4}  
\usepackage{graphicx}  
\usepackage{dcolumn}   
\usepackage{bm}        
\usepackage{amssymb}   

\newcommand{\btwb}     {\mbox{${\cal B}(t \to Wb)/{\cal B}(t \to Wq)$}}
\newcommand{\Btwb}     {\mbox{$\mathbf{{\cal B}(t \to Wb)/{\cal B}(t \to Wq)}$}}
\newcommand{\met}{\not{\!\!\!E_T}}
\newcommand{\ttbar}     {\mbox{$t\bar{t}$}}

\newcommand{\ntt}{\mbox{$N_{t\bar{t}}$}}
\newcommand{\ttb}{\mbox{$t\bar{t}$}}
\newcommand{\mtop}	{\mbox{$m_t$}}
\newcommand{\pntag}     {P_{n_{\mathrm{tag}}}}
\begin{document}

\hspace{5.2in} \mbox{Fermilab-Pub-04/06-037-E}

\title{Measurement of $\Btwb$ at $\mathbf{\sqrt{s}}$ = 1.96 TeV}
%
\author{                                                                      
V.M.~Abazov,$^{36}$                                                           
B.~Abbott,$^{75}$                                                             
M.~Abolins,$^{65}$                                                            
B.S.~Acharya,$^{29}$                                                          
M.~Adams,$^{52}$                                                              
T.~Adams,$^{50}$                                                              
M.~Agelou,$^{18}$                                                             
J.-L.~Agram,$^{19}$                                                           
S.H.~Ahn,$^{31}$                                                              
M.~Ahsan,$^{59}$
% ...
\\                                                                            
\vskip 0.30cm                                                                 
\centerline{(D\O\ Collaboration)}                                             
\vskip 0.30cm                                                                 
}                                                                             
\affiliation{                                                                 
\centerline{$^{1}$Universidad de Buenos Aires, Buenos Aires, Argentina}       
\centerline{$^{2}$LAFEX, Centro Brasileiro de Pesquisas F{\'\i}sicas,         
                  Rio de Janeiro, Brazil}                                     
\centerline{$^{3}$Universidade do Estado do Rio de Janeiro,                   
                  Rio de Janeiro, Brazil}                                     
\centerline{$^{4}$Instituto de F\'{\i}sica Te\'orica, Universidade            
                  Estadual Paulista, S\~ao Paulo, Brazil}                     
\centerline{$^{5}$University of Alberta, Edmonton, Alberta, Canada,           
               Simon Fraser University, Burnaby, British Columbia, Canada,}   
\centerline{York University, Toronto, Ontario, Canada, and                    
         McGill University, Montreal, Quebec, Canada}                         
\centerline{$^{6}$Institute of High Energy Physics, Beijing,                  
                  People's Republic of China}                                 
\centerline{$^{7}$University of Science and Technology of China, Hefei,       
                  People's Republic of China}                                 
\centerline{$^{8}$Universidad de los Andes, Bogot\'{a}, Colombia}             
\centerline{$^{9}$Center for Particle Physics, Charles University,            
                  Prague, Czech Republic}                                     
\centerline{$^{10}$Czech Technical University, Prague, Czech Republic}        
\centerline{$^{11}$Center for Particle Physics, Institute of Physics,         
                   Academy of Sciences of the Czech Republic,                 
                   Prague, Czech Republic}                                    
\centerline{$^{12}$Universidad San Francisco de Quito, Quito, Ecuador}        
\centerline{$^{13}$Laboratoire de Physique Corpusculaire, IN2P3-CNRS,         
                  Universit\'e Blaise Pascal, Clermont-Ferrand, France}       
\centerline{$^{14}$Laboratoire de Physique Subatomique et de Cosmologie,      
                  IN2P3-CNRS, Universite de Grenoble 1, Grenoble, France}     
\centerline{$^{15}$CPPM, IN2P3-CNRS, Universit\'e de la M\'editerran\'ee,     
                  Marseille, France}                                          
\centerline{$^{16}$IN2P3-CNRS, Laboratoire de l'Acc\'el\'erateur              
                  Lin\'eaire, Orsay, France}                                  
\centerline{$^{17}$LPNHE, IN2P3-CNRS, Universit\'es Paris VI and VII,         
                  Paris, France}                                              
\centerline{$^{18}$DAPNIA/Service de Physique des Particules, CEA, Saclay,    
                  France}                                                     
\centerline{$^{19}$IReS, IN2P3-CNRS, Universit\'e Louis Pasteur, Strasbourg,  
                France, and Universit\'e de Haute Alsace, Mulhouse, France}   
\centerline{$^{20}$Institut de Physique Nucl\'eaire de Lyon, IN2P3-CNRS,      
                   Universit\'e Claude Bernard, Villeurbanne, France}         
\centerline{$^{21}$III. Physikalisches Institut A, RWTH Aachen,               
                   Aachen, Germany}                                           
\centerline{$^{22}$Physikalisches Institut, Universit{\"a}t Bonn,             
                  Bonn, Germany}                                              
\centerline{$^{23}$Physikalisches Institut, Universit{\"a}t Freiburg,         
                  Freiburg, Germany}                                          
\centerline{$^{24}$Institut f{\"u}r Physik, Universit{\"a}t Mainz,            
                  Mainz, Germany}                                             
\centerline{$^{25}$Ludwig-Maximilians-Universit{\"a}t M{\"u}nchen,            
                   M{\"u}nchen, Germany}                                      
\centerline{$^{26}$Fachbereich Physik, University of Wuppertal,               
                   Wuppertal, Germany}                                        
\centerline{$^{27}$Panjab University, Chandigarh, India}                      
\centerline{$^{28}$Delhi University, Delhi, India}                            
\centerline{$^{29}$Tata Institute of Fundamental Research, Mumbai, India}     
\centerline{$^{30}$University College Dublin, Dublin, Ireland}                
\centerline{$^{31}$Korea Detector Laboratory, Korea University,               
                   Seoul, Korea}                                              
\centerline{$^{32}$SungKyunKwan University, Suwon, Korea}                     
\centerline{$^{33}$CINVESTAV, Mexico City, Mexico}                            
\centerline{$^{34}$FOM-Institute NIKHEF and University of                     
                  Amsterdam/NIKHEF, Amsterdam, The Netherlands}               
\centerline{$^{35}$Radboud University Nijmegen/NIKHEF, Nijmegen, The          
                  Netherlands}                                                
\centerline{$^{36}$Joint Institute for Nuclear Research, Dubna, Russia}       
\centerline{$^{37}$Institute for Theoretical and Experimental Physics,        
                  Moscow, Russia}                                             
\centerline{$^{38}$Moscow State University, Moscow, Russia}                   
\centerline{$^{39}$Institute for High Energy Physics, Protvino, Russia}       
\centerline{$^{40}$Petersburg Nuclear Physics Institute,                      
                   St. Petersburg, Russia}                                    
\centerline{$^{41}$Lund University, Lund, Sweden, Royal Institute of          
                   Technology and Stockholm University, Stockholm,            
                   Sweden, and}                                               
\centerline{Uppsala University, Uppsala, Sweden}                              
\centerline{$^{42}$Physik Institut der Universit{\"a}t Z{\"u}rich,            
                    Z{\"u}rich, Switzerland}                                  
\centerline{$^{43}$Lancaster University, Lancaster, United Kingdom}           
\centerline{$^{44}$Imperial College, London, United Kingdom}                  
\centerline{$^{45}$University of Manchester, Manchester, United Kingdom}      
\centerline{$^{46}$University of Arizona, Tucson, Arizona 85721, USA}         
\centerline{$^{47}$Lawrence Berkeley National Laboratory and University of    
                  California, Berkeley, California 94720, USA}                
\centerline{$^{48}$California State University, Fresno, California 93740, USA}
\centerline{$^{49}$University of California, Riverside, California 92521, USA}
\centerline{$^{50}$Florida State University, Tallahassee, Florida 32306, USA} 
\centerline{$^{51}$Fermi National Accelerator Laboratory, Batavia,            
                   Illinois 60510, USA}                                       
\centerline{$^{52}$University of Illinois at Chicago, Chicago,                
                   Illinois 60607, USA}                                       
\centerline{$^{53}$Northern Illinois University, DeKalb, Illinois 60115, USA} 
\centerline{$^{54}$Northwestern University, Evanston, Illinois 60208, USA}    
\centerline{$^{55}$Indiana University, Bloomington, Indiana 47405, USA}       
\centerline{$^{56}$University of Notre Dame, Notre Dame, Indiana 46556, USA}  
\centerline{$^{57}$Iowa State University, Ames, Iowa 50011, USA}              
\centerline{$^{58}$University of Kansas, Lawrence, Kansas 66045, USA}         
\centerline{$^{59}$Kansas State University, Manhattan, Kansas 66506, USA}     
\centerline{$^{60}$Louisiana Tech University, Ruston, Louisiana 71272, USA}   
\centerline{$^{61}$University of Maryland, College Park, Maryland 20742, USA} 
\centerline{$^{62}$Boston University, Boston, Massachusetts 02215, USA}       
\centerline{$^{63}$Northeastern University, Boston, Massachusetts 02115, USA} 
\centerline{$^{64}$University of Michigan, Ann Arbor, Michigan 48109, USA}    
\centerline{$^{65}$Michigan State University, East Lansing, Michigan 48824,   
                   USA}                                                       
\centerline{$^{66}$University of Mississippi, University, Mississippi 38677,  
                   USA}                                                       
\centerline{$^{67}$University of Nebraska, Lincoln, Nebraska 68588, USA}      
\centerline{$^{68}$Princeton University, Princeton, New Jersey 08544, USA}    
\centerline{$^{69}$State University of New York, Buffalo, New York 14260, USA}
\centerline{$^{70}$Columbia University, New York, New York 10027, USA}        
\centerline{$^{71}$University of Rochester, Rochester, New York 14627, USA}   
\centerline{$^{72}$State University of New York, Stony Brook,                 
                   New York 11794, USA}                                       
\centerline{$^{73}$Brookhaven National Laboratory, Upton, New York 11973, USA}
\centerline{$^{74}$Langston University, Langston, Oklahoma 73050, USA}        
\centerline{$^{75}$University of Oklahoma, Norman, Oklahoma 73019, USA}       
\centerline{$^{76}$Oklahoma State University, Stillwater, Oklahoma 74078, USA}
\centerline{$^{77}$Brown University, Providence, Rhode Island 02912, USA}     
\centerline{$^{78}$University of Texas, Arlington, Texas 76019, USA}          
\centerline{$^{79}$Southern Methodist University, Dallas, Texas 75275, USA}   
\centerline{$^{80}$Rice University, Houston, Texas 77005, USA}                
\centerline{$^{81}$University of Virginia, Charlottesville, Virginia 22901,   
                   USA}                                                       
\centerline{$^{82}$University of Washington, Seattle, Washington 98195, USA}  
}                                                                             
\date{February 27, 2006}

\begin{abstract} 
We present the measurement of $R = \btwb$ in $p\bar{p}$ collisions at 
$\sqrt{s}=1.96$~TeV, using 230~pb$^{-1}$ of data collected by the D\O\  
experiment at the Fermilab Tevatron Collider. We fit simultaneously
$R$ and the number ($\ntt$) of selected top quark pairs
($t\bar{t}$),
to the number of identified $b$-quark jets 
in events with one electron or one muon, three or more jets, and 
high transverse energy imbalance.
To improve sensitivity, kinematical properties of events with no 
identified $b$-quark jets are included in the fit. We measure 
$R = 1.03^{+0.19}_{-0.17}~\rm(stat+syst) $,  in good agreement with the standard model.
We set lower limits of $R > 0.61$ and $|V_{tb}| > 0.78$ at 95\% confidence level.
\end{abstract}

\maketitle 

Within the standard model (SM), the top quark decays 
99.8\%  of the time to a $W$ boson and a $b$ quark, 
with the ratio $R $= $\btwb$ 
(here $q$ refers to $d, s$, or $b$ quarks) expressible in terms of the 
Cabbibo-Kobayashi-Maskawa (CKM)
matrix elements~\cite{ckm}  \( R= \frac{\mid V_{tb}\mid^2} 
{\mid V_{tb}\mid^2 + \mid V_{ts}\mid^2 + \mid V_{td}\mid^2} \). 
The unitarity of the CKM matrix and experimental constraints on its elements~\cite{PDG}  
yield the SM prediction $0.9980 < R < 0.9984$
at the 90\% C.L. 
Nevertheless, a fourth generation of quarks or non-SM processes 
in the production or decay of the top quark could lead to significant 
deviations from the SM. So far, measurements of $R$ by the CDF 
collaboration~\cite{CDFref,CDFref2} have not established a deviation 
of $R$ from unity.

In the present analysis, we assume that the top quark decays into
a $W$ boson, but that the associated quark can be $d$, $s$, or $b$. 
Lepton + jets final states arise in $\ttb$ when one $W$ boson decays
leptonically and the other into  a $q\bar{q}'$ pair. About 6\% 
of the signal arises from $\ttbar$ events in which both $W$ bosons 
decay leptonically, but one charged
lepton is not reconstructed, while additional jets are produced by
initial or final state radiation. 
In this Letter, we report the measurement of $R$ 
in the lepton (electron or muon) + jets channel 
($\ell$ + jets). The lepton can come either
from a direct $W$ decay or from $W \to \tau \to e/\mu$.
We use $b$-jet identification ($b$-tagging) techniques,
exploiting the long lifetime of $B$ hadrons, to separate 
$\ttb$ events from the background processes.
The data were collected by the D\O\ experiment from August 2002 through March 2004,  
and correspond to an integrated luminosity of $230$~pb${}^{-1}$. 

The D\O\ detector incorporates a tracking system, calorimeters, and a muon
spectrometer~\cite{d0det}. The tracking system is made up of a silicon \
micro-strip tracker (SMT) and a central fiber tracker (CFT), located inside a 2~T
superconducting solenoid. The tracking system provides  efficient charged
particle detection in the pseudorapidity region $|\eta| < 3$~\cite{eta}. The
SMT strip  pitch of 50--80~$\mu$m allows a precise determination of the
primary interaction vertex (PV) and an accurate measurement of the impact
parameter of a track relative to the PV~\cite{ip}. These are key
components of the lifetime-based $b$-tagging algorithms. The PV
is required to be within the fiducial region of the SMT and to contain at 
least three tracks. The calorimeter
consists of a barrel section covering $|\eta|<1.1$, and two end-caps 
extending the coverage to  $|\eta|\approx4.2$.  The muon spectrometer
surrounds the calorimeter and consists of three layers of drift chambers 
and several layers of scintillators~\cite{muon_detector}. A 1.8 T iron
toroidal magnet is located outside the innermost layer of the muon system.
The luminosity is calculated from the rate of {\mbox{$p\bar p$}}\ inelastic 
collisions, detected by two arrays of scintillation counters mounted 
close to the beam-pipe on the front surfaces of the calorimeter end-caps.

We select data in the electron and muon decay channels
by requiring an isolated electron with $p_T>20$~GeV and
$|\eta|<1.1$, or an isolated muon with $p_T>20$~GeV and $|\eta|<2.0$. 
The lepton isolation criteria are based on calorimeter and tracking
information. More details on lepton identification and trigger requirements are 
available in Ref.~\cite{ljetstopo_paper}. In both channels, we require 
the missing transverse energy ($\met$) to exceed $20$~GeV and not 
be colinear with the direction of the lepton projected on the transverse 
plane. The candidate events must be accompanied by jets with $p_T>15$~GeV 
and rapidity $|y|<2.5$~\cite{eta}. 
Jets are defined using a cone algorithm with radius  
$\Delta{\cal R}=0.5$~\cite{jet}.

We use a secondary vertex tagging (SVT) algorithm to reconstruct
displaced vertices produced by the decay of $B$ hadrons inside jets.
Secondary vertices are reconstructed from two or more tracks
satisfying:
$p_T>1$ GeV, $\geq 1$ hits in the SMT detector, and impact parameter 
significance $d_{ca}/\delta_{d_{ca}}>3.5$~\cite{ip}. Tracks identified as arising 
from $K^0_S$ or $\Lambda$ decays or from $\gamma$ conversions 
are not used. If the secondary vertex reconstructed within a 
jet has a decay-length significance $L_{xy}/\delta_{L_{xy}}>7$~\cite{dl}, 
the jet is defined as $b$-tagged.  Events with exactly 1 ($\geq 2$) 
$b$-tagged jets are referred to as 1-tag (2-tag) events. 
Events with no $b$-tagged jets are referred to as 0-tag events. 
A prediction for the number of background events and the fractions of 
\ttbar\ events in the 0, 1, and 2-tag samples require the
probabilities for different types of jets ($b$-, $c$-, and light-quark
jets) to be $b$-tagged. The calculation of these
probabilities is presented in Ref.~\cite{d0_top_btag}. We fit
simultaneously $R$ and the total number of $\ttbar$ events in the 0,
1, and 2-tag samples ($\ntt$) to the number of observed
1-tag and 2-tag events, and, in 0-tag events,
to the shape of a discriminant variable ${\cal D}$ that exploits kinematic
differences between the backgrounds and the $\ttbar$ signal.

The main background in this analysis is from the production of
leptonically decaying $W$ bosons produced in association with jets
($W$+jets).
Most of the jets accompanying the $W$ boson
originate from $u$, $d$, and $s$ quarks and gluons ($W$+light jets). 
Between $2$\% and $14$\% of  $W$+jets events contain heavy-flavor
jets, arising from gluon splitting into 
$b\bar{b}$ or $c\bar{c}$ ($Wb\bar{b}$ or $Wc\bar{c}$, respectively).
About $5$\% of the $W$+jets events contain a single $c$ quark
that originates from $W$-boson radiation from an $s$ quark in
the proton or anti-proton sea ($s \to Wc$). A sizable background arises from 
strong production of two or more jets (``multijets''), with one of the jets 
misidentified as an isolated lepton, and accompanied by large 
$\met$ resulting from mismeasurement of jet energies.
Significantly smaller contributions to the selected sample arise from
$Z$+jets, $WW$, $WZ$, $ZZ$, and single top quark production. 
Together, these five smaller backgrounds are expected to contribute 
from 1\% to 7\% of the selected sample, depending on the number of 
$b$-tagged jets, and are referred to below as ``other'' backgrounds. 

Normalization of the backgrounds begins with the determination 
of the number of multijet events in the selected sample. The multijet 
background is determined using control data samples and probabilities 
for jets to mimic isolated lepton signatures, also derived from 
data~\cite{ljetstopo_paper}. Subtracting this background also provides 
the fraction of events with a truly isolated high-$p_T$ lepton (i.e. $\ttbar$ 
and all backgrounds, except multijets).
The contributions from single top quark, 
$Z$+jets, and diboson production are determined from Monte Carlo
simulation (MC). The remainder corresponds either to $\ttbar$ or $W$+jet production. 
The signal and background processes are generated using {\sc alpgen}  \cite{alpgen} 
with \mtop\ $=175$ GeV. {\sc pythia} \cite{pythia} is used for fragmentation and 
decay. $B$ hadron decays are modeled via {\sc evtgen} 
\cite{evtgen}. A full detector simulation is performed using 
{\sc geant} \cite{geant}.

In an analysis based on the SM, with $R$ $\approx$ 1, the $\ttbar$ event tagging 
probabilities are computed assuming that each of the signal
events contains two $b$-jets~\cite{d0_top_btag}. In the present
analysis, the top quark can also decay into a light quark 
($d$ or $s$) and a $W$ boson. The ratio $R$ determines the 
fraction of $\ttbar$ events with 0, 1, and 2 $b$-jets and therefore 
how $\ttbar$ events are distributed among
the 0, 1, and 2-tag samples. In order to derive the $\ttb$ 
event tagging probability as a function of $R$, we determine the
tagging probability for the three following scenarios ({\it i})
$t\bar{t} \to W^+b~W^-\bar{b}$ (to be referred to as $tt \to
bb$), ({\it ii}) $t\bar{t} \to W^+b~W^-\bar{q}_l$ or its charge
conjugate (referred to as $tt \to bq_l$), and ({\it iii})
$t\bar{t} \to W^+q_l~W^-\bar{q}_l$ (referred to as $tt \to q_l q_l$),
where $q_l$ denotes either a $d$ or $s$ quark. The probabilities 
$P_{{n}_{\rm tag}}$ to observe $n_{\rm tag}$ = 0, 1, 
or $\geq$ 2 $b$-tagged jets are computed separately for the 
three types of $\ttb$ events, using the probabilities for
each type of jet ($b$, $c$, or light-quark jet) to be $b$-tagged. The probabilities 
$P_{n_{\rm tag}}$  in the three scenarios are then combined 
to obtain the $\ttb$ tagging probability as a function of $R$,
\( \pntag (tt) = R^2 \pntag (tt \to bb) + 
  2R(1-R) \pntag (tt \to b q_l) +
  (1-R)^2 \pntag (tt \to q_l q_l)
\), where the subscript $n_{\rm tag}$ runs over 0, 1, and $\geq$ 2 tags.
Table~\ref{tab:prediction_vs_observation} compares the observed
number of events in the 0, 1, and 2-tag samples with the sum of the
predicted backgrounds and the fitted number of $\ttbar$ events.

The fraction of $\ttbar$ events in the $\ell$ + $\geq$ 4 jets ($\ell$ +
3 jets) 0-tag 
sample changes from 10\% (2\%) for $R=1$ to 22\% (4\%) for $R=0$. The size of this
contribution is of the order of the Poisson uncertainty on the number 
of events in the 0-tag sample. Therefore the
number of observed 0-tag events is a poor constraint on $R$ and 
$\ntt$. We achieve a tighter constraint on the number of $\ttbar$ events in the 
0-tag sample by constructing a discriminant function ${\cal D}$ for
0-tag events in the $\ell$ + $\geq$4 jets sample, that combines kinematical event
properties to discriminate between $\ttbar$ signal and $W$+jets 
background. The signal to background ratio in the $\ell$ + 3
jets, 0-tag sample is five times smaller than in the corresponding 
$\geq$ 4 jets sample. Therefore we do not consider such a discriminant for
$\ell$ + 3 jets, 0-tag events. We select four variables that provide 
good discrimination between signal and background and that are well modeled 
by the MC.  The discriminant function is built from: 
({\it i}) the event sphericity {$\mathcal S$}, constructed from the 
four-momenta of the jets, ({\it ii}) 
the event centrality ${\mathcal C}$, defined as the ratio 
of the scalar sum of the $p_T$ of the jets to the 
scalar sum of the energies of the jets,
({\it iii}) $K'_{T{\rm min}} = \Delta
{\cal R}_{jj}^{\rm min}p_T^{\rm min}/E_T^W$, where $\Delta
{\cal R}_{jj}^{\rm min}$ is the minimum separation in $\eta-\phi$
space between pairs of jets, $p_T^{\rm min}$ is the $p_T$ of the
lower-$p_T$ jet of that pair, and $E_T^W$ is the scalar sum of the lepton
transverse momentum and~${\mbox{$\not\!\!E_T$}}$, 
 and ({\it iv}) $H^{'}_{T2}$ = $H_{T2}/H_z$, 
where $H_{T2}$ is the scalar sum of the $E_T$ for all jets 
excluding the leading jet and $H_z$ is the scalar sum of the 
absolute value of the momenta of all the jets, the lepton and 
the neutrino along the $z$-direction~\cite{neutrinopz}.
Sphericity and centrality characterize the event shape and are 
described in Ref.~\cite{tensor}. In order to reduce the 
dependence on modeling of soft radiation and the underlying event, 
only the four highest-$p_T$ jets are used 
to determine these variables. 

The discriminant function is constructed using the method described in 
Ref.~\cite{topmass}. Neglecting correlations among the input variables
$x_1,x_2,...$, the discriminant function can be approximated by the expression:  
\begin{eqnarray}
\label{eq:discr1}
{\mathcal D} &=& \frac{\prod_i s_i(x_i)/b_i(x_i)}{\prod_i s_i(x_i)/b_i(x_i) + 1} \;,
\end{eqnarray}
where $s_i(x_i)$ and $b_i(x_i)$ are the normalized distributions of variable $x_i$ 
for signal and background, respectively.
As constructed, the discriminant peaks near zero for background, and near 
one for signal. The shapes of the discriminant for $\ttbar$ 
and $W$+jets events are derived from MC.

The shape of the discriminant for the multijet background
is obtained from a control data sample, selected 
by requiring that the lepton candidates fail 
the isolation criteria. The other backgrounds ($Z$+jets, 
diboson, and single top quark) have discriminant 
distributions close to those of the $W$+jet 
events, and contribute to 1\% of the 0-tag sample. In the 
final fit, we assume that these processes have the same 
discriminants as the $W$+jets events. The background 
normalization in the $\ell + \geq 4$ jets, 0-tag sample is 
extracted from the discriminant fit rather than from MC.
To verify that the kinematic variables used in the 
discriminant are well modeled by the simulation we compare
data and MC distributions in two control samples.
To avoid biasing the measurement with respect 
to $R$, we choose control samples where $b$-tagging is not applied, and to avoid bias with
respect to $\ntt$ we select events with little $\ttbar$ content: $\ell$ + 2 jets and 
$\ell$ + 3 jets. In $\ell$ + 2 jets events, the fraction of $\ttbar$ events is negligible, 
whereas it makes up about 5\% of the $\ell$ + 3 jets events.

\begin{figure*}[tbh]
\center{\includegraphics[width=0.99\textwidth]{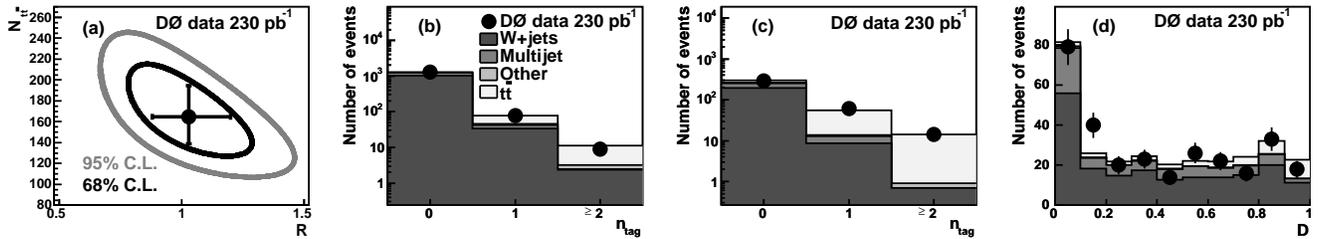}}
\caption{\label{fig:control_l3j} {(a) The 68\% and 95\%
    statistical confidence contours in the ($R$, $\ntt$) plane. 
    The point indicates the best fit to data. 
    Observed number of events and fitted sample composition in 
    the 0, 1, and 2-tag samples 
    (b) in the $\ell$ + 3 jets sample and (c) in the
    $\ell$ + $\geq$4 jets sample. (d) Observed
    and fitted distribution of the discriminant ${\cal D}$.
}}
\end{figure*}

In order to measure $R$ and $\ntt$, we perform a binned maximum likelihood fit.
The data are binned in thirty bins:
({\it i})   twenty bins of the discriminant ${\cal D}$ in the $e$ +
$\geq$4 jets and $\mu$ + $\geq$4 jets, 0-tag samples,
({\it ii}) two bins for the two 0-tag samples in $e$ + 3 jets and $\mu$ + 3 jets,
({\it iii})  four bins for the four 1-tag samples (electron or muon and
3 or 4 jets), and
({\it iv})   four bins for the four 2-tag samples (electron or muon and 3 or 4 jets).
In each bin, we predict the number of events that corresponds to 
the sum of the expected background and signal. The signal
contribution is a function of $R$ and $\ntt$. To predict the number of events in
each bin of the discriminant ${\cal D}$, we use its expected
distribution for $W$+jets background
and $\ttbar$ signal. As described earlier, the normalization of the
multijet background is estimated by counting events in orthogonal
control samples. Statistical fluctuations in the 
number of events in the control samples are taken into account.
We incorporate systematic 
uncertainties into the likelihood by using nuisance parameters~\cite{nuisance_parameters}.
All preselection efficiencies, tagging probabilities, and shapes of the discriminant
${\cal D}$ are functions of the nuisance parameters. The likelihood
contains one Gaussian term for each nuisance parameter. The value of
$R$ that maximizes the total likelihood is \( R = 
1.03^{+0.19}_{-0.17}~\rm(stat+syst) \), in good agreement with the SM
expectation. A summary of statistical and systematic uncertainties is
given in Table~\ref{tab:systematics_R}. The fit also yields the total
number of $\ttbar$ events in the 0, 1, and 2-tag samples,
\(  \ntt = 163^{+29}_{-27}~\rm(stat) \).
The result of the two-dimensional fit is shown in the ($R$, $\ntt$) plane 
in Fig.~\ref{fig:control_l3j}(a), with the 68\% and 95\% 
contours of statistical confidence. In Fig.~\ref{fig:control_l3j}(b) and 
Fig.~\ref{fig:control_l3j}(c), we compare the observed 
number of events to the sum of the predicted backgrounds and the
fitted $\ttbar$ contribution, in the 0, 
1 and 2-tag samples for events with 3 jets and $\geq$ 4 jets.
In Fig.~\ref{fig:control_l3j}(d), we compare the observed distribution
of the discriminant ${\cal D}$ with the corresponding distribution for
the sum of the predicted backgrounds and the fitted $\ttbar$ contribution.

We extract lower limits on $R$ and the CKM matrix element $|V_{tb}|$ 
assuming $|V_{tb}| = \sqrt{R}$. Using a Bayesian approach with the
prior $\pi(R) = 1$ for $0 \leq  R \leq 1$ and $\pi(R) = 0$ otherwise,
we obtain $R > 0.78$ at the 68\% C.L. and $R > 0.61$ at the 95\% C.L. For the CKM
matrix element $|V_{tb}|$, we obtain $|V_{tb}| > 0.88$ at 68\% C.L., 
and $|V_{tb}| > 0.78$ at the 95\% C.L.

In summary, we performed the most accurate measurement of $R$ to date, 
 \( R = 1.03^{+0.19}_{-0.17}~\rm(stat+syst) \), in good agreement with
the SM.

%
%

\begin{center}
\begin{table}
\caption{\label{tab:prediction_vs_observation}{\sl Observed number of
    events, predicted backgrounds and fitted $\ntt$.}}
\begin{tabular}{lr@{$\pm$}lr@{$\pm$}lr@{$\pm$}l}
\hline
\hline
$\ell$ + 3 jets & \multicolumn{2}{c}{0-tag}   & \multicolumn{2}{c}{1-tag}   & \multicolumn{2}{c}{$\geq$ 2-tag}  \\ 
\hline
         W+jets    & 1032 & 38    & 34   & 5   &  2.4 & 0.4 \\
      Multi-jet    & 192  & 23    &  8.3 & 1.5 &  \multicolumn{2}{c}{$\! 0.1 ^{+0.3}_{-0.1}$} \\
      Other bkg    & 18.4 & 1.3   &  4.3 & 0.3 &  0.7 & 0.1 \\
\hline
Fitted $\ttbar$    & 32.4 & 1.6   & 32.3 & 1.6 &  8.2 & 0.5 \\
\hline 
Total              & 1275 & 44    & 79   & 5   & 11.4 & 0.8 \\
\hline
Observed           & \multicolumn{2}{c}{1277} &  \multicolumn{2}{c}{79} & \multicolumn{2}{c}{9}  \\ 
\hline
\hline
$\ell$ + $\geq$ 4 jets & \multicolumn{2}{c}{0-tag}   &
\multicolumn{2}{c}{1-tag}   & \multicolumn{2}{c}{$\geq$ 2-tag}  \\ 
\hline
         W+jets    & 193  & 17  &  8.8 & 1.2 &  0.7 & 0.1 \\
      Multi-jet    & 65   &  9  &  4.1 & 1.1 &  0.0 & 0.4 \\
      Other bkg    &  2.9 & 0.4 &  1.2 & 0.2 &  0.2 & 0.1 \\
\hline
Fitted $\ttbar$    & 35.6 & 2.8 & 41.5 & 3.3 & 13.5 & 1.4 \\
\hline
 Total             &  297 &  19 & 56   &  4  & 14.4 & 1.4 \\
\hline
Observed    &  \multicolumn{2}{c}{291} & \multicolumn{2}{c}{62} &   \multicolumn{2}{c}{14}  \\ 
\hline
\hline
\end{tabular}
\end{table}
\end{center}

%
\begin{center}
  \begin{table}
  \caption{\label{tab:systematics_R}{Summary of statistical and systematic uncertainties on $R$.}}
    \begin{tabular}{lc}
      \hline
      \hline
      \multicolumn{2}{p{7cm}}{Uncertainties on $R$} \\
      \hline
      Statistical                           & +0.17 $-$0.15  \\
      $b$-tagging efficiency                & +0.06 $-$0.05  \\ 
      Background modeling                   & +0.05 $-$0.04  \\ 
      Jet identification and energy calibration & +0.04 $-$0.03  \\ 
      Multijet background                  & $\pm$0.02      \\ 
      \hline
      Total error                           & +0.19 $-$0.17 \\
      \hline
      \hline
    \end{tabular}
  \end{table}
\end{center}
%

%
%

%
We thank the staffs at Fermilab and collaborating institutions, 
and acknowledge support from the 
DOE and NSF (USA);
CEA and CNRS/IN2P3 (France);
FASI, Rosatom and RFBR (Russia);
CAPES, CNPq, FAPERJ, FAPESP and FUNDUNESP (Brazil);
DAE and DST (India);
Colciencias (Colombia);
CONACyT (Mexico);
KRF and KOSEF (Korea);
CONICET and UBACyT (Argentina);
FOM (The Netherlands);
PPARC (United Kingdom);
MSMT (Czech Republic);
CRC Program, CFI, NSERC and WestGrid Project (Canada);
BMBF and DFG (Germany);
SFI (Ireland);
The Swedish Research Council (Sweden);
Research Corporation;
Alexander von Humboldt Foundation;
and the Marie Curie Program.
%


\begin{thebibliography}{99}
%
\bibitem[*]{kurca}
On leave from IEP SAS Kosice, Slovakia.
\bibitem[\#]{parashar}
Visitor from Purdue University Calumet, Hammond, Indiana, USA.
\bibitem[\dag]{voutilainen}
Visitor from Helsinki Institute of Physics, Helsinki, Finland.
%
\vskip 0.25cm
  
  \bibitem{ckm} N. Cabbibo, Phys. Rev. Lett. {\bf 10}, 531 (1963);
    M. Kobayashi and T. Maskawa, Prog. Theor. Phys. {\bf 49}, 652 (1973).

  \bibitem{PDG}
    S. Eidelman {\it et al.}, Phys. Lett. B {\bf 592}, 1 (2004).
      
  \bibitem{CDFref}
    CDF Collaboration, T. Affolder {\it et al.}, Phys. Rev. Lett. {\bf 86}, 3233 (2001).

  \bibitem{CDFref2}
    CDF Collaboration, T. Affolder {\it et al.}, Phys. Rev. Lett. {\bf 95}, 102002 (2005).

  \bibitem{d0det}
    D\O\ Collaboration, V.~Abazov {\it et al.}, ``The Upgraded D\O\ Detector,"
    submitted to Nucl. Instr. and Methods in Phys. Res. A, hep-physics/0507191.

  \bibitem{eta} Rapidity $y$ and pseudorapidity $\eta$ are defined as functions
    of the parameter $\beta$ and polar angle $\theta$ w.r.t. the proton beam line, 
    as $y(\theta,\beta) \equiv
    {1 \over 2} \ln{[(1+\beta\cos{\theta})/(1-\beta\cos{\theta})]}$ and
    $\eta(\theta) \equiv y(\theta,1)$, where $\beta$ is the ratio of a particle's 
    momentum to its energy.

  \bibitem{ip} Impact parameter is defined as the distance of closest 
    approach ($d_{ca}$) of the track to the primary vertex in the 
    plane transverse to the beam line. Impact parameter significance is defined as
    $d_{ca}/\delta_{d_{ca}}$, where $\delta_{d_{ca}}$ is the error on $d_{ca}$.

  \bibitem{muon_detector} 
    V.~Abazov {\it et al.}, Nucl. Instrum. Methods in Phys. Res. A {\bf 552}, 372-398 (2001).

  \bibitem{ljetstopo_paper} 
    D\O\ Collaboration, V. Abazov {\it et al.}, Phys. Lett. B {\bf 626}, 45 (2005).

  \bibitem{jet} We use the iterative, seed-based cone algorithm 
    including midpoints, as described on p. 47 in G.~C.~Blazey {\it
      et al.}, in Proceedings of the Workshop: {\it ``QCD and Weak Boson
      Physics in Run~II,''}  edited by U.~Baur, R. K.~Ellis, and
    D. Zeppenfeld, FERMILAB-PUB-00-297 (2000).

  \bibitem{dl} Decay length $L_{xy}$ is defined as the distance 
    from the primary to the secondary vertex in the plane transverse to the beam line.
    Decay length significance is defined as $L_{xy}/\delta_{L_{xy}}$, where
    $\delta_{L_{xy}}$ is the uncertainty on $L_{xy}$.

  \bibitem{alpgen} 
    M.\ L.~Mangano {\it et al.}, JHEP {\bf 07}, 001 (2003).
    
  \bibitem{pythia}
    T.\ Sj\"ostrand {\it et al.}, Comput. Phys. Commun. {\bf 135}, 238 (2001).
    
  \bibitem{evtgen} D.\ J.\ Lange, Nucl. Instrum. Methods in Phys. Res. A {\bf 462},
    152 (2001).
    
  \bibitem{geant}
    R.\ Brun and F.\ Carminati, CERN Program Library Long Writeup W5013, 1993 (unpublished).

  \bibitem{d0_top_btag}
    D\O\ Collaboration, V. Abazov {\it et al.}, Phys. Lett. B {\bf 626}, 35 (2005).
    
  \bibitem{neutrinopz} The neutrino momentum along the $z$ direction
    ($p^{\nu}_z$) is determined by assuming that each event contains one $W$
    boson. The $W$ mass is used as an additional constraint to derive $p^{\nu}_z$.
    The solution with the smallest $|p^{\nu}_z|$ is chosen.

  \bibitem{tensor} 
    V.\ Barger, J.\ Ohnemus, and R.J.N.~Phillips, Phys. Rev. D {\bf 48}, R3953 (1993).

  \bibitem{topmass} 
    D\O\ Collaboration, B.\ Abbott {\it et al.}, Phys. Rev. D {\bf 58}, 052001 (1998).

  \bibitem{nuisance_parameters}
    E.T. Jaynes, ``Probability Theory,'' University Press, Cambridge, 2003.

\end{thebibliography}
\end{document}